\documentclass[12pt]{article}
\usepackage{amsfonts}
\usepackage{latexsym,amsmath}
\usepackage{amssymb,array,graphicx,xcolor}
\begin{document}
\title{The inverse xgamma distribution: statistical properties and different methods of estimation}
\begin{scriptsize}
\author{Abhimanyu Singh Yadav $^{a}$, Sudhansu S. Maiti $^{b}$, Mahendra Saha $^{a}$\footnote{Corresponding author. e-mail: mahendrasaha@curaj.ac.in}\\ 
and Arvind Pandey $^{a}$\\
\small $^{a}$Department of Statistics, Central University of Rajasthan, Rajasthan, India\\
\small $^{b}$ Department of Statistics, Visva-Bharati University, Santiniketan, India}
\end{scriptsize}
\date{}
\maketitle
\begin{abstract}
This paper proposed a new probability distribution named as inverse xgamma distribution (IXGD). Different mathematical and statistical properties,viz., reliability characteristics, moments, inverse moments, stochastic ordering and order statistics of the proposed distribution have been derived and discussed. The estimation of the parameter of IXGD has been approached by different methods of estimation, namely, maximum likelihood method of estimation (MLE), Least square method of estimation (LSE), Weighted least square method of estimation (WLSE), Cram\`er-von-Mises method of estimation (CME) and maximum product spacing method of estimation (MPSE). Asymptotic confidence interval (ACI) of the parameter is also obtained. A simulation study has been carried out to compare the performance of  the obtained estimators and corresponding ACI in terms of average widths and corresponding coverage probabilities. Finally, two real data sets have been used to demonstrate the applicability of IXGD in real life situations. 
\end{abstract}
\section{Introduction.}
Sen et al. ($2016$) introduced a finite mixture of exponential ($\theta$) and gamma ($3,\theta$) distributions with mixing proportion $\pi_1=\theta/(1+\theta)$ and $\pi_2=1-\pi_1$, where $\pi_1,\pi_2$ denote the mixing proportions that are non-negative and sum to one, to obtained a probability distribution, named as xgamma distribution (XGD). The probability density function (PDF) and cumulative distribution function(CDF) of the XGD are, respectively, given by 
\begin{eqnarray}\label{eq1}
f(y;\theta)&=&\frac{\theta^2}{(1+\theta)}\left(1+\frac{\theta}{2}.y^2\right)e^{-\theta y}~~;y>0,~\theta>0
\\&=&0~~~~~~~~~~~~~~~~~~~~~~~~~~~~~~~~~~;\mbox{otherwise}\nonumber.
\end{eqnarray}
\begin{eqnarray}\label{eq2}
F(y;\theta)&=&1-\frac{\left(1+\theta+\theta y+\frac{\theta^2 y^2}{2}\right)}{(1+\theta)}e^{-\theta y}~~;y>0,~\theta>0
\\&=&0~~~~~~~~~~~~~~~~~~~~~~~~~~~~~~~~~~~~;\mbox{otherwise}\nonumber.
\end{eqnarray}
Sen et al. ($2016$) investigated mathematical, structural and survival properties of the XGD and they have found that in many cases the XGD has more flexibility than the exponential distribution. A new probability distribution, namely, inverse xgamma distribution (IXGD) is introduced in this article. The inverse of the XGD is considered in order to obtain the form of the inverse xgamma distribution and hence the name proposed.\\\\

The objective of this article is two fold: First, we introduced a new probability distribution and studied the several statistical properties of IXGD, as the inverted version of XGD, introduced by Sen et al. ($2016$). Second, different methods of estimation have been employed to estimate the unknown parameter of IXGD. Further, $95\%$ asymptotic confidence interval (ACI) of the parameter based on MPSE has been constructed. To the best of our knowledge thus far, no attempt has been made to introduce the inverted version of XGD. Our aim is to fill up this gap through this present study.\\

The rest of the article is organized as follows: In Section $2$, we have developed a new probability distribution, called IXGD. In Section $3$, we have studied different statistical properties and related measures of IXGD. Different methods of estimation of the parameter of IXGD have been considered in Section $4$. In Section $5$, ACI of the parameter of IXGD has been obtained. Monte Carlo simulation has been carried out to see the performance of the estimates of the parameters in mean squared error sense. Empirical applications are presented and discussed in Section $6$. Finally, concluding remarks are given in Section $7$.
\section{The inverse xgamma distribution.}
In recent years, the researcher has proposed number of methods to introduce a new probability distribution. The inverse transfromation method of baseline variables is one of them and the resulting distribution is parsimonious in parameter, for example inverse exponential distribution (IED) [see, Keller and Kamath ($1982$)], inverse Rayleigh distribution (IRD) [see, Voda ($1972$)], inverse lindley distribution (ILD) [see, Sharma et al. ($2015$)] etc. The same approach has been used to introduce the inverted form of XGD. If a random variable $Y$ has XGD ($\theta$) with PDF given in ($\ref{eq1}$), then the random variable $X=(1/Y)$ is said to follow the inverse xgamma distribution (IXGD) with PDF is of the following form:
\begin{eqnarray}\label{eq3}
f(x)&=&\frac{\theta^2}{(1+\theta)}.\frac{1}{x^2}\left(1+\frac{\theta}{2}.\frac{1}{x^2}\right)e^{-\theta/x}~~;x>0,~\theta>0
\\&=&0~~~~~~~~~~~~~~~~~~~~~~~~~~~~~~~~~~;\mbox{otherwise}\nonumber.
\end{eqnarray}
It is denoted by $X \sim \mbox{IXGD}(\theta)$. The CDF of IXGD is given by
\begin{eqnarray}\label{eq4}
F(x)&=&\left(1+\frac{\theta}{(1+\theta)}.\frac{1}{x}+\frac{\theta^2}{2(1+\theta)}.\frac{1}{x^2}\right)e^{-\theta/x}~~;x>0,~\theta>0
\end{eqnarray}
The shape of the density and distribution functions for IXGD are presented in Figure 1. 
\begin{figure}
	\includegraphics[width=6.5in, height=3.5in]{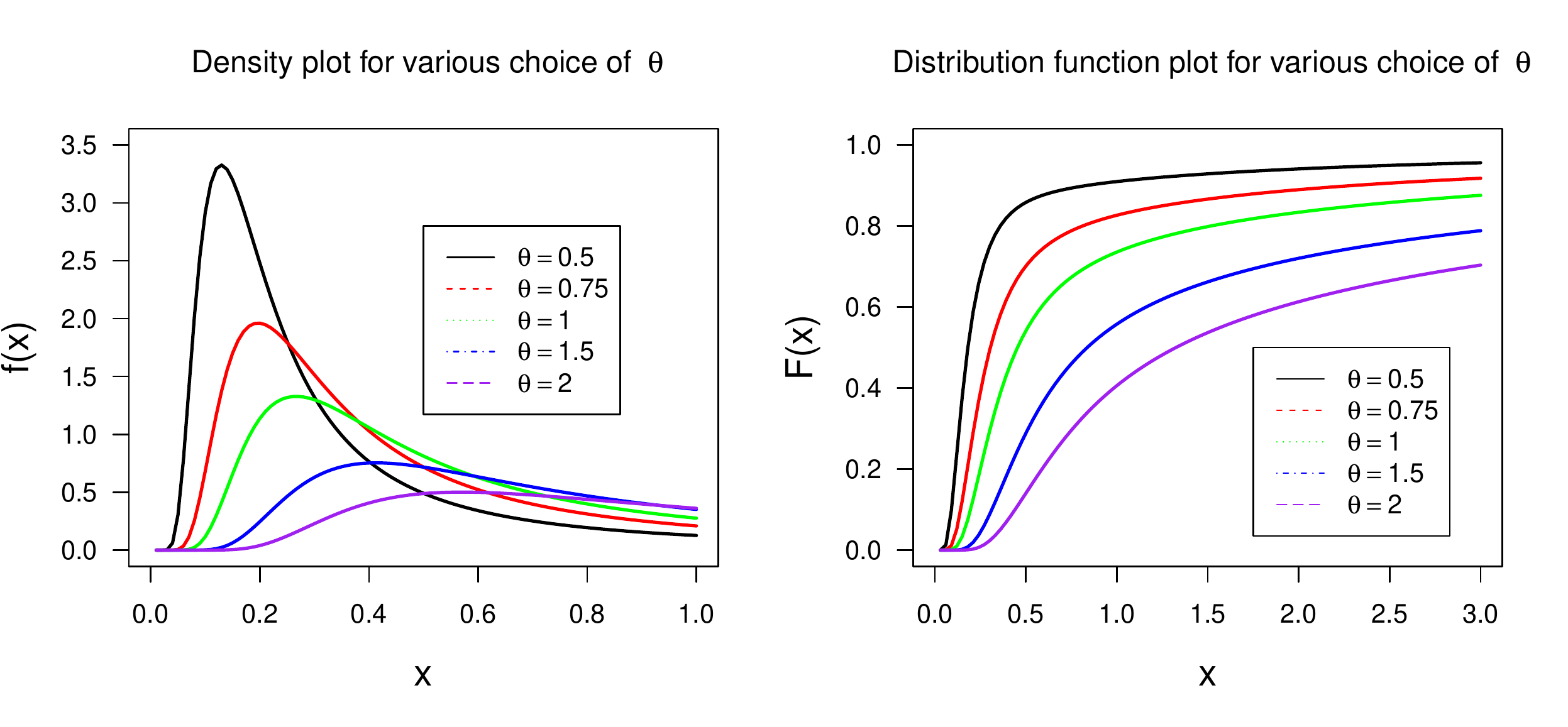}
	\caption{Density and distribution function plot}
\end{figure}
\section{Some statistical properties.}
In the following subsections, different statistical properties, viz., reliability characteristics, moments and inverse-moments, stochastic ordering and order statistics have been discussed.
\subsection{Reliability Characteristics:}
The basic tools for studying the ageing and associated characteristics of any lifetime equipments are the reliability and hazard functions. The reliability and the hazard function of IXGD($\theta$) are given below:  
\begin{itemize}
	\item The reliability function $R(t)$, which is the probability that an item not failing prior to some time $t(\ge 0)$, is defined by
	\begin{eqnarray}
		R(t)&=&\nonumber P[X \geq t]\\&=&\nonumber 1-F(t)\\&=&1-\left\{\left(1+\frac{\theta}{(1+\theta)}.\frac{1}{x}+\frac{\theta^2}{2(1+\theta)}.\frac{1}{x^2}\right)e^{-\theta/x}\right\}.
	\end{eqnarray}
\item The hazard rate function (or failure rate function) for a continuous distribution with PDF $f(t)$, CDF $F(t)$, and survival function (SF) $S(t)$ is the conditional probability of failure, given it has survived up to time $t(\ge 0)$ and is defined as
\begin{eqnarray*}
	H(t)&=&\frac{f(t)}{1-F(t)}\\&=&\frac{f(t)}{S(t)}.
\end{eqnarray*}
For the IXGD, the hazard rate function is given by
\begin{eqnarray}
	H(t)&=& \frac{\left\{\frac{\theta^2}{(1+\theta)}.\frac{1}{t^2}\left(1+\frac{\theta}{2}.\frac{1}{t^2}\right)e^{-\theta/t}\right\}}{\left[1-\left\{\left(1+\frac{\theta}{(1+\theta)}.\frac{1}{t}+\frac{\theta^2}{2(1+\theta)}.\frac{1}{t^2}\right)e^{-\theta/t}\right\}\right]}.
\end{eqnarray}
\item The reverse hazard rate function (or reverse failure rate function) for a continuous distribution with PDF $f(t)$, CDF $F(t)$ is defined as
\begin{eqnarray}
	h(t)&=&\nonumber \frac{f(t)}{F(t)}\\&=&\frac{\theta^2(2t^2+\theta)}{(2t^2+2\theta t^2+2\theta t+\theta^2)}.\frac{1}{t^2}.
\end{eqnarray}
\end{itemize}
The shape of the reliability and hazard functions for IXGD are presented in Figure $2$. From Figure $2$, it is clear that the proposed distribution accommodate the shape of non-monotone failure rate pattern. Such pattern of failure rate is very obvious in clinical trial studies and in reliability studies, thus, IXGD can be an alternative choice to analyse such data set. 
\begin{figure}
	\includegraphics[width=6.5in, height=3.5in]{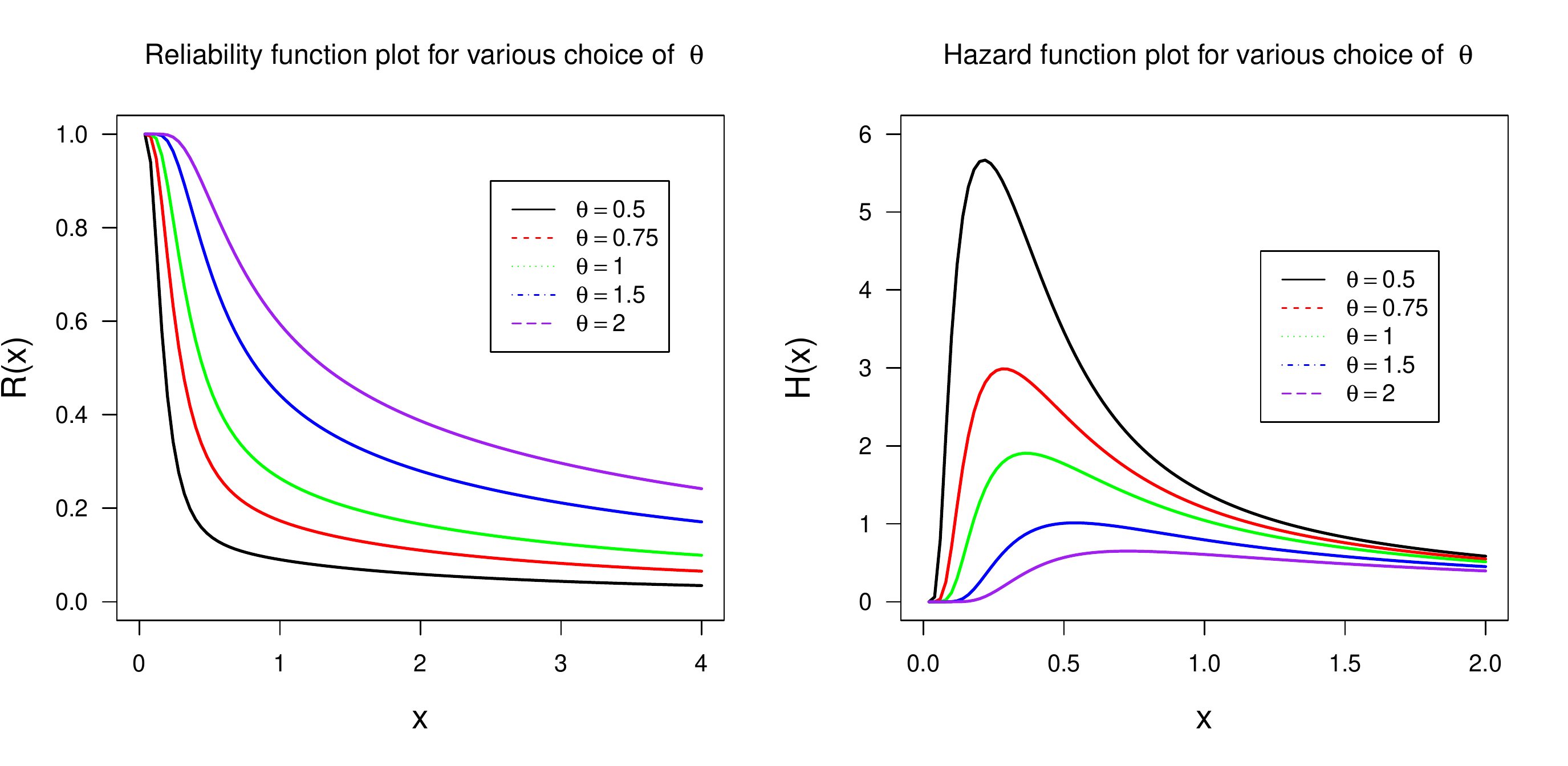}
	\caption{Reliability and hazard function plot}
\end{figure}
\subsection{Moments and related measures:}
The $r$-th order moment about origin of IXGD is given by
\begin{eqnarray*}
\mu^{'}_{r}&=&E\left(X^r\right)
               \\&=&\int\limits_{0}^{\infty}x^r.\frac{\theta^2}{(1+\theta)}.\frac{1}{x^2}\left(1+\frac{\theta}{2}.\frac{1}{x^2}\right)e^{-\theta/x}.dx
               \\&=&\left\{\frac{\theta^{r+1}}{(1+\theta)}\Gamma(1-r)+\frac{\theta^{r}}{2(1+\theta)}\Gamma(3-r)\right\}~;~~r<1.
\end{eqnarray*}
The above expression indicate that moment of IXGD will exist only when $r<1$. Therefore, the evaluation of inverse moments may be of interest. The $r$-th order inverse moment about origin of IXGD is given by
\begin{eqnarray}
\mu^{'}_{r^{-1}}&=&E\nonumber \left(\frac{1}{X^r}\right)
               \\&=&\nonumber \int\limits_{0}^{\infty}\frac{1}{x^r}.\frac{\theta^2}{(1+\theta)}.\frac{1}{x^2}\left(1+\frac{\theta}{2}.\frac{1}{x^2}\right)e^{-\theta/x}.dx
               \\&=&\frac{\theta^2}{2(1+\theta)}\left[ \frac{2\Gamma(r+1)}{\theta^{r+1}}+\frac{\Gamma(r+3)}{\theta^{r+2}}\right] ~~~~;r=1, 2, 3~ \&~ 4.
\end{eqnarray}
\subsection{Harmonic mean and other moments:}
The harmonic mean for the density function as expressed in ($\ref{eq3}$) is obtained by 
\begin{equation}
E\left( \frac{1}{x}\right) =\int_{x} \frac{1}{x}f(x,\theta) dx.
\end{equation}
The above equation can also be calculated from the expression of inverse of moment by putting $r=1$. Hence, after simplification we get
\begin{eqnarray}
E\left( \frac{1}{x}\right) =\dfrac{\theta+3}{\theta(1+\theta)}.
\end{eqnarray}
\subsection{Stochastic Ordering:}
Stochastic ordering of a positive random variable is a very important property to study the comparative behaviour of a random variable. Recall some basic definitions:
A random variable X is said to be smaller than a random variable Y in the
\begin{itemize}
	\item stochastic order $(X\le_{st} Y)$ if $F_X(x) \ge F_Y(x)$ for all $x$
	\item  hazard rate order $(X\le_{hr} Y)$ if $h_X(x) \ge h_Y(x)$ for all $x$
	\item mean residual life order $(X\le_{mrl} Y)$ if $m_X(x) \le m_Y(x)$ for all $x$
	\item likelihood ratio order $(X\le_{lr} Y)$ if $f_X(x)/f_Y(x)$ decreases in $x$
\end{itemize}
The following implications based on this property are illustrated by Shaked \& Shanthikumar (1994). 
$$X\le_{lr}Y\Rightarrow X\le_{hr}Y\Rightarrow X\le_{mrl}Y$$ and hence $$X\le_{hr}Y\Rightarrow X\le_{st} Y$$
The following theorem shows that IXGD is ordered with respect to the strongest likelihood ratio ordering.\\
\textbf{Theorem:} Let $X \sim IXGD(\theta_1)$ and $Y \sim IXGD(\theta_2)$. If $\theta_1 > \theta_2$, then $X\le_{lr} Y$ and hence it implies oredering in others also. 

\textbf{Proof:} \\
\begin{eqnarray*}
\phi(x)&=&\frac{f_{\theta_1}(x)}{f_{\theta_2}(x)}\\&=&\frac{\theta^2_1(1+\theta_2)}{\theta^2_2(1+\theta_1)}\left\{\left(\frac{2x^2+\theta_1}{2x^2+\theta_2}\right)e^{-\frac{1}{x}(\theta_1-\theta_2)} \right\}
\end{eqnarray*}
Taking logarithm both sides, we can write
\begin{eqnarray*}
\ln \phi(x)&=&\ln K+\ln(2x^2+\theta_1)-\ln(2x^2+\theta_2)-\frac{1}{x}(\theta_1-\theta_2)
\end{eqnarray*}
Taking partial derivative both sides, we have
\begin{eqnarray*}
\frac{\partial \ln \phi(x)}{\partial x}&=&\frac{1}{\phi(x)}.\phi^{'}(x)\\&=&\left\{ \frac{4x}{(2x^2+\theta_1)(2x^2+\theta_2)}(\theta_2-\theta_1)-\frac{(\theta_2-\theta_1)}{x^2}\right\},
\end{eqnarray*}
that implies
\begin{equation*}
\phi^{'}(x)=\phi(x)(\theta_2-\theta_1)\left\{\frac{4x}{(2x^2+\theta_1)(2x^2+\theta_2)}-\frac{1}{x^2}\right\}
\end{equation*} 
Now, if $\theta_1 > \theta_2$, then $\phi^{'}(x)< 0$. Hence, $\phi(x)$ decreases in $x$ and this implies $X \leq_{st} Y$.
\subsection{Order Statistics:}
Let $X_{(1)},~X_{(2)},~...,~X_{(n)}$ are the $n$ ordered random sample observed from density function (\ref{eq3}). Then, the distribution of $r$-th order statistic is obtained by using the following expressions as follows:
\begin{eqnarray}\label{eq5}
f_r(x)&=&\frac{1}{\beta(r,n-r+1)}.\sum\limits_{k=0}^{n-r}(-1)^k{{n-r}\choose k}F^{r+k-1}(x).f(x)
\end{eqnarray}
and the $r$-th order CDF $F_r(x)$ is 
\begin{eqnarray}\label{eq6}
F_r(x)&=&\sum\limits_{j=r}^{n}\sum\limits_{k=0}^{n-j}{{n}\choose j}{{n-j}\choose k}(-1)^kF^{j+k}(x)
\end{eqnarray}
Hence, using Equations (\ref{eq3}), (\ref{eq4}) in (\ref{eq5}), the PDF and the CDF of $r$th order statistics are, respectively, given by
\begin{eqnarray}\label{eq7}
f_r(x)&=&\nonumber\frac{1}{\beta(r,n-r+1)}.\frac{\theta^2}{(1+\theta)}\sum\limits_{k=0}^{n-r}(-1)^k{{n-r}\choose k}\left(1+\frac{\theta}{(1+\theta)}.\frac{1}{x}+\frac{\theta^2}{2(1+\theta)}.\frac{1}{x^2}\right)^{r+k-1} \times \\&&
\left(\frac{1}{x^2}\left(1+\frac{\theta}{2}.\frac{1}{x^2}\right)\right)e^{-\theta(r+k)/x}
\end{eqnarray}
\begin{eqnarray}\label{eq8}
F_r(x)&=&\nonumber\sum\limits_{j=r}^{n}\sum\limits_{k=0}^{n-j}{{n}\choose j}{{n-j}\choose k}(-1)^k\left(1+\frac{\theta}{(1+\theta)}.\frac{1}{x}+\frac{\theta^2}{2(1+\theta)}\frac{1}{x^2}\right)^{j+k}  \times \\&& e^{-\theta(j+k)/x}.
\end{eqnarray}
The distributions (PDF \& CDF) of the smallest and the largest order statistics in case of IXGD are obtained by putting $r=1$ and $r=n$ in Equations ($\ref{eq7}$) and ($\ref{eq8}$) respectively.
\section{Different methods of estimation of parameter $\theta$.}
In this section, we have used five methods of estimation to estimate the unknown parameter $\theta$, namely, maximum likelihood method of estimation (MLE), least squares method of estimation (LSE), weighted least squares method of estimation (WLSE), Cram\`er-von-Mises estimator method estimation (CME) and maximum product of spacings method of estimation (MPSE) respectively.\\

\subsection{Maximum likelihood method of estimation:}
Let $(x_1,~x_2,~...,~x_n)$ be the random sample of size $n$ drawn from the IXGD, given in Equation ($\ref{eq3}$). The maximum likelihood estimator (MLE) of $\theta$ for given $x$ is obtained as follows:\\
The likelihood function of $\theta$ is given by 
\begin{eqnarray*}
L(\theta \mid x)&=&\prod\limits_{i=1}^{n}\left\{\frac{\theta^2}{(1+\theta)}.\frac{1}{x_i^2}\left(1+\frac{\theta}{2}.\frac{1}{x_i^2}\right)e^{-\theta/x_i}  \right\}.
\end{eqnarray*}
Taking logarithm both sides, the log-likelihood function is given by
\begin{eqnarray*}
l(\theta \mid x)&=&\ln L(\theta \mid x)\\&=&2n \ln \theta-n\ln (1+\theta)-2\sum\limits_{i=1}^{n} \ln x_i + \sum\limits_{i=1}^{n} \ln \left(1+\frac{\theta}{2 x^2_i}\right)-\theta \sum\limits_{i=1}^{n}\frac{1}{x_i}
\end{eqnarray*}
The resulting partial derivative of the log-likelihood function
\begin{equation}
\frac{\partial l(\theta \mid x)}{\partial\theta}=\sum\limits_{i=1}^{n}\left\{\frac{1}{2 x^2_i+\theta}\right\}+\frac{2n}{\theta}-\frac{n}{(1+\theta)}-\sum\limits_{i=1}^{n}\frac{1}{x_i}=0
\end{equation}
yield  the MLE of $\theta$. Equating this partial derivative to zero does not yield closed-form solution for MLE of $\theta$ and thus a numerical method is used for solving this equation.
\subsection{Ordinary and weighted least square methods of estimation:}
The least square estimator (LSE) and the weighted least square estimator (WLSE) were proposed by Swain et al. ($1988$) to estimate the parameters of Beta distributions. Suppose $F(x_{(j)})$ denotes the distribution function of the ordered random variables $x_{(1)}<x_{(2)}<\cdots <x_{(n)}$, where, $\{x_{1},x_{2},\cdots ,x_{n}\}$ is a random sample of size $n$ from a distribution function $F(\cdot)$. Therefore, in this case, the LSE of $\theta$, say, $\hat{\theta}_{LSE}$ can be obtained by minimizing
$$
 S(\theta) =\sum_{i=1}^{n}\left[ F\left(x_{i:n}| \theta \right) -\frac{i}{n+1}\right] ^{2}
$$
with respect to $\theta$, where, $F(\cdot)$ is the CDF, given in Equation (\ref{eq4}). Equivalently, it can be obtained by solving:
\begin{eqnarray}
&&\displaystyle\sum_{i=1}^{n}\left[ F\left( x_{i:n}\mid \theta\right) -\frac{i}{n+1}\right] \eta _{1}\left( x_{i:n}\mid \theta\right) =0,
\end{eqnarray}
where,
\begin{equation}\label{eq10} 
\eta_{1}\left( x_{i:n}\mid \theta \right)=-\frac{\theta e^{-\theta/x_i}}{2x_i^3(1+\theta)^2}\left(4x_i^2+2\theta x_i^2+\theta x_i + \theta + \theta^2\right) 
\end{equation}
The WLSE of $\theta$, say, $\hat{\theta}_{WLSE}$, can be obtained by minimizing
\begin{eqnarray}
\displaystyle W\left( \kappa, \lambda \right) = \sum_{i=1}^{n} \frac {\left(n+1\right)^{2}\left( n+2\right)}{i\left( n-i+1\right)} \left[ F\left(
x_{i:n}\mid \theta \right) - \frac {i}{n+1} \right]^{2}.
\end{eqnarray}
This estimator can also be obtained by solving:
\begin{eqnarray}
&& \displaystyle \sum_{i=1}^{n}\frac {\left( n+1\right)^{2}\left( n+2\right)%
}{i\left( n-i+1\right)} \left[ F\left( x_{i:n}\mid \theta \right) -
\frac {i}{n+1} \right] \eta_{1}\left( x_{i:n}\mid \theta\right)=0,
\end{eqnarray}
 where, $\eta_{1}\left( x_{i:n}\mid \theta \right)$ is defined in ($\ref{eq10}$). 
\subsection{Cram\`er-von-Mises method of estimation:}
To motivate our choice of Cramer-von Mises type minimum distance estimators, MacDonald ($1971$) provided empirical evidence that the bias of the estimator is smaller than the other minimum distance estimators. Thus, The Cramer-von Mises estimators of $\theta$, say, $\hat{\theta}_{CME}$ can be obtained by minimizing the function 
\begin{eqnarray}
C(\theta) =\frac{1}{12n}+\sum_{i=1}^{n}\left[ F\left(x_{i:n} \mid \theta \right) -\frac{2i-1}{2n}\right] ^{2}
\end{eqnarray}
with respect to $\theta$. The estimator can also be obtained by solving the non linear equation
\begin{eqnarray}
&&\displaystyle\sum_{i=1}^{n}\left[ F\left( x_{i:n}\mid \theta\right) -\frac{2i-1}{2n}\right] \eta _{1}\left( x_{i:n}\mid \theta\right)=0,
\end{eqnarray}
 where, $\eta_{1}\left( x_{i:n}\mid \theta \right)$ is defined in ($\ref{eq10}$).
\subsection{Maximum product of spacings method of estimation:}
The maximum product spacing method has been introduced by Cheng and Amin ($1979,\\~1983$) as an alternative to MLE for the estimation of the unknown parameters of continuous univariate distributions. This method was also derived independently by Ranneby ($1984$) as an approximation to the Kullback-Leibler measure of information. To motivate our choice, Cheng and Amin ($1983$) proved that this method is as efficient as the MLE and consistent under more general conditions. Let us define 
\begin{equation}
D_{i}(\theta)=F\left( x_{i : n}\mid \theta \right)
-F\left( x_{ i-1 : n}\mid \theta \right), \qquad i=1,2,\ldots ,n,
\end{equation}
where, $F(x_{ 0 : n}\mid \theta)=0$ and $F( x_{ n+1 : n}\mid \theta)=1.$ Clearly, \ $\sum_{i=1}^{n+1} D_i (\theta) =1.$. The MPSE $\hat{\theta}_{MPS}$, of the parameter $\theta$ are obtained by maximizing the geometric mean of the spacings with respect to $\theta$, given as
\begin{equation}
G\left(\theta \right) =\left[ \prod\limits_{i=1}^{n+1}D_{i}(\theta)\right] ^{\frac{1}{n+1}},
\end{equation}
or, equivalently, by maximizing the function 
\begin{equation}
H\left(\theta \right) =\frac{1}{n+1}\sum_{i=1}^{n+1}\log D_{i}(\theta).
\end{equation}
The estimator $\hat{\theta}_{MPS}$ of the parameter $\theta$ can be obtained by solving the
non-linear equation
\begin{eqnarray}
\frac{\partial }{\partial \theta }H\left(\theta \right)=\frac{1}{n+1}
\sum\limits_{i=1}^{n+1}\frac{1}{D_{i}(\theta)} \left[ \eta_1(x_{i:n} |  \theta) - \eta_1 (x_{i-1:n} |  \theta)\right] &=&0
\end{eqnarray}
where, $\eta _{1}\left( \cdot \mid \theta \right) $ is given by ($\ref{eq10}$). 
\section{Asymptotic confidence interval of $\theta$.}
Here, we consider the asymptotic confidence interval (ACI) based on MPSE, as MPSE of $\theta$ performed better in mean squared error (MSE) sense among the other estimates (MLE, LSE, WLSE and CME). Cheng and Amin ($1979$), Ghosh and Jammalamadaka ($2001$) already mentioned and shows that the maximum product spacing method also shows asymptotic properties as MLE. Keeping this in mind, we have considered the Fisher information and is obtained as;
\begin{equation}
I(\hat{\theta})=-E\left[ \dfrac{\partial^{2} \ln L(\theta)}{\partial\theta^{2}}\right]  
\end{equation}
where, $$\dfrac{\partial^{2} \ln L(\theta)}{\partial\theta^{2}}=-\sum\limits_{i=1}^{n}\left\{\frac{1}{(2 x^2_i+\theta)^2}\right\}-\frac{2n}{\theta^2}+\frac{n}{(1+\theta)^2}.$$
Therefore, the asymptotic variance $\sigma^2_{\theta\theta}$ of $\theta$ is obtained as
$$
\sigma^2_{\theta\theta}=\left[\dfrac{1}{I(\hat{\theta})}\right] _{MPSE}
$$
The $100 (1-\alpha)$\% ACI based on MPSE of $\theta$ is given by
$$
\left\{\hat{\theta}_{MPSE}\mp Z_{(\alpha/2)}\sqrt{\left( \sigma^{2}_{\theta\theta}\right)}\right\},
$$
where, $Z_{(\alpha/2)}$ is the upper $(\alpha/2)$-th point of the standard normal distribution.\\\\
Also, to study the confidence interval (CI), we have considered the estimated average widths and coverage probabilities of ACI using MSE of $\theta$. In addition, the average widths of ACIs are calculated based on the $B=1,000$ different trials. The average widths and corresponding coverage probabilities are given by
$$
\mbox{Average width}=\frac{\sum\limits_{i=1}^{B}\left(U_i-L_i\right)}{B},
$$
and
$$
\mbox{Coverage probability}=\frac{\mbox{Number}\left(L_W \leq \theta \leq U_p\right)}{B},
$$
where, $L_W$ and $U_P$ denote the $100(1-\alpha)\%$ CIs based on $B$ replicates.
\section{Simulation and Discussion.}
In this section, we have carried out a Monte Carlo simulation study to assess the performance of the proposed estimators (MLE, LSE, WLSE, CME and MPSE) of the parameter $\theta$ for IXGD, discussed in Section $3$. In particular, we have considered the different variations of sample sizes $n=10,~20,~50,~100$ and the parameter values $\theta=0.1,~0.5,~1.0,~1.5$ respectively. For each design, sample with each of size $n$ are drawn from the original sample and replicated $5,000$ times. First, we have calculated the average estimates (AV) of the parameter $\theta$ using MLE, LSE, WLSE, MPSE and CME along with the corresponding MSEs. The results are reported in Table $1$.\\ 
\begin{eqnarray*}
AV=\frac{1}{5000}\sum_{j=1}^{5000}\theta_j
\end{eqnarray*}
\begin{eqnarray*}
MSE=\frac{1}{5000}\sum_{j=1}^{5000}(\theta_j-\theta)^2.
\end{eqnarray*}
From Table $1$, it has been observed that as the sample sizes increases, the MSEs of all estimators are decreases. It verifies the consistency of all the estimators that we have considered. It has been also observed that the MSEs of MPSE of $\theta$ are less for all the considered choices of $n$ and $\theta$. Further, the interval estimation of the parameter is also considered and corresponding results are reported in Table $2$. Table $2$ showed the estimated average widths and coverage probabilities of $95\%$ ACI of the parameter $\theta$ for IXGD using MPSE. Here also, it has been observed that as the sample sizes increases, the average widths decreases. All simulations were performed using programs written in the open source statistical package $R$ [see, Ihaka and Gentleman, ($1996$)].\\
\begin{table}[ht]
\textbf{Table 1: True value of $\theta$ along with average estimates and corresponding MSEs for IXGD.}\\
\begin{center}
\begin{tabular}{|l|l|l|l|l|l|l|l|l|l|l|l|l|}\hline
\multicolumn{1}{|c|}{}&
\multicolumn{1}{|c|}{}&
\multicolumn{5}{|c|}{Estimates of $\theta$ and corresponding MSEs} \\
\cline{3-7}
$n$  & $\theta$   &  MLE       &    LSE     &   WLSE     &  CME       &   MPSE      \\
\hline
$10$ & $0.10$     & $0.102854$ & $0.101512$ & $0.101270$ & $0.102144$ & $0.097421$ \\
     &            & $0.000385$ & $0.000457$ & $0.000439$ & $0.000457$ & $0.000344$ \\ 
\hline
$20$ & $0.10$     & $0.101592$ & $0.101217$ & $0.101025$ & $0.101560$ & $0.098079$  \\
     &            & $0.000197$ & $0.000228$ & $0.000218$ & $0.000229$ & $0.000183$  \\    
\hline
$50$ & $0.10$     & $0.100781$ & $0.100311$ & $0.100357$ & $0.100455$ & $0.098992$  \\
     &            & $0.000072$ & $0.000083$ & $0.000077$ & $0.000083$ & $0.000069$  \\    
\hline
$100$& $0.10$     & $0.100284$ & $0.100091$ & $0.100151$ & $0.100164$ & $0.099238$  \\
     &            & $0.000035$ & $0.000040$ & $0.000038$ & $0.000040$ & $0.000035$  \\
\hline
\hline
$10$ & $0.50$     & $0.521981$ & $0.751896$ & $2.324485$ & $0.789168$ & $0.487656$ \\
     &            & $0.0130953$ & $1.782553$ & $191.100200$ & $2.254722$ & $0.011006$ \\
\hline
$20$ & $0.50$     & $0.513173$ & $0.718121$ & $2.284905$ & $0.774732$ & $0.491954$  \\
     &            & $0.005711$ & $1.626593$ & $190.9996$ & $1.518383$ & $0.005195$  \\ 
\hline
$50$ & $0.50$     & $0.506318$ & $0.527461$ & $1.978845$ & $0.528769$ & $0.495569$  \\
     &            & $0.002262$ & $0.179746$ & $78.79531$ & $0.182699$ & $0.002158$  \\ 
\hline
$100$& $0.50$     & $0.500999$ & $0.504581$ & $1.37044$ & $0.505124$ & $0.494758$  \\
     &            & $0.001028$ & $0.027757$ & $76.54193$ & $0.027899$ & $0.001002$ \\   
\hline
\hline
$10$ & $1.00$     & $1.067974$ & $1.072587$ & $5.07754$ & $1.185889$ & $0.986976$ \\
     &            & $0.064716$ & $2.146959$ & $241.826$ & $2.146005$ & $0.053416$ \\  
\hline
$20$ & $1.00$     & $1.028478$ & $1.185968$ & $4.71047$ & $1.095143$ & $0.979043$  \\
     &            & $0.026007$ & $1.264714$ & $166.107$ & $1.312496$ & $0.022954$  \\
\hline
$50$ & $1.00$     & $1.010089$ & $1.237631$ & $3.19992$ & $1.084109$ & $0.985321$  \\
     &            & $0.0091631$ & $1.07522$ & $121.167$ & $1.17262$ & $0.008626$  \\ 
\hline
$100$& $1.00$     & $1.004157$ & $1.426787$ & $2.11971 $ & $1.08019$ & $0.989713$  \\
     &            & $0.004993$ & $0.832140$ & $111.24400$ & $1.106002$ & $0.004813$  \\
\hline
\hline
$10$ & $1.50$     & $1.61502$ & $1.570825$ & $5.144353$ & $1.185889$ & $0.986976$ \\
     &            & $0.1655082$ & $0.251852$ & $241.826234$ & $2.146005$ & $0.053416$ \\  
\hline
$20$ & $1.50$     & $1.564556$ & $1.554483$ & $1.75837$ & $1.564219 $ & $1.482625$  \\
     &            & $0.073945$ & $0.1023667$ & $153.753451$ & $1.103020$ & $0.044771$  \\
\hline
$50$ & $1.50$     & $1.514692$ & $1.520526$ & $1.637223$ & $1.524409$ & $1.473679$  \\
     &            & $0.025980$ & $0.101587$ & $76.340366$ & $0.811026$ & $0.024301$ \\ 
\hline
$100$& $1.50$     & $1.507803$ & $1.413624$ & $1.468612$ & $1.453052$ & $1.483952$  \\
     &            & $0.011712$ & $0.080194$ & $14.224941$ & $0.278361$ & $0.011236$  \\
\hline
\end{tabular}
\end{center}
\label{tab1}
\end{table}
  \begin{table}
 \textbf{Table 2: $\theta$ and its estimated average widths and coverage probabilities of ACI for IXGD.}\\
 \begin{center}
 \begin{tabular}{|l|l|l|l|l|l|l|l|l|l|l|l|l|}\hline
 \multicolumn{1}{|c|}{}&
 \multicolumn{1}{|c|}{}&
 \multicolumn{2}{|c|}{Confidence Limits}&
 \multicolumn{1}{|c|}{}&
 \multicolumn{1}{|c|}{}\\
 \cline{3-4}
 $n$     & $\theta$  &    $L$     &   $U$    &  Average   & Cov. \\
         &           &            &         &  Width      & Prob. \\
 \hline    
 $10$    & $0.10$    & $0.0000$ & $0.2170$  & $0.2170$  & $0.939$ \\
 \hline
 $20$    & $0.10$    & $0.0000$ & $0.2166$  & $0.2166$  & $0.941$ \\
 \hline 
 $50$    & $0.10$    & $0.0000$ & $0.2164$  & $0.2164$  & $0.941$ \\
 \hline
 $100$   & $0.10$    & $0.0000$ & $0.2143$  & $0.2143$  & $0.943$ \\
 \hline
 \hline
  $10$    & $0.50$    & $0.0000$ & $1.1510$  & $1.1510$  & $0.938$ \\ 
   \hline
   $20$    & $0.50$   & $0.0000$ & $1.1418$  & $1.1418$  & $0.941$ \\
   \hline 
   $50$    & $0.50$   & $0.0000$ & $1.1334$  & $1.1334$  & $0.941$ \\
   \hline
   $100$   & $0.50$   & $0.0000$ & $1.1166$  & $1.1166$  & $0.944$ \\
   \hline 
   \hline
  $10$    & $1.00$    & $0.0000$ & $2.4052$  & $2.4052$  & $0.939$ \\ 
   \hline
 $20$    & $1.00$    & $0.0000$ & $2.3219$  & $2.3219$  & $0.940$ \\ 
   \hline
 $50$    & $1.00$    & $0.0000$ & $2.3094$  & $2.3094$  & $0.941$ \\ 
 \hline
  $100$    & $1.00$    & $0.0000$ & $2.2732$  & $2.2732$  & $0.942$ \\ 
 \hline
 \hline 
  $10$    & $1.50$    & $0.0000$ & $3.5353$  & $3.5353$  & $0.938$ \\ 
  \hline
  $20$    & $1.50$    & $0.0000$ & $3.4840$  & $3.4840$  & $0.940$ \\
  \hline 
  $50$    & $1.50$    & $0.0000$ & $3.4460$  & $3.4460$  & $0.941$ \\
  \hline
  $100$   & $1.50$    & $0.0000$ & $3.3722$  & $3.3722$  & $0.942$ \\
  \hline 
 \end{tabular}
 \end{center}
 \label{tab2}
 \end{table}
\section{Real Life Examples.}
Here, we consider two real data sets to show the practical applicability of the proposed model. We check whether the considered data sets actually come from the IXGD or not by goodness of fit test. For this purpose, we compared the newly introduced model IXGD with well known one parameter inverted family of distributions, namely, inverse exponential distribution (IED), inverse Rayleigh distribution (IRD), inverse Lindley distribution (ILD). This procedure is based on the Kolmogorov-Smirnov (K-S) statistic and it compares an empirical and a theoretical model by computing the maximum absolute difference between the empirical and theoretical distribution functions and is defined as $D_{n}= Sup_{x}|F_{n}(x) -F(x; \alpha)|$, where $Sup_{x}$ is the supremum of the set of distances, $F_{n}(x)$ is the empirical distribution function and $F(x; \alpha)$ is the CDF. Note that, K-S statistic to be used only to verify the goodness-of-fit and not as a discrimination criteria. Therefore, we consider four discrimination criteria based on the log-likelihood function evaluated at the maximum likelihood estimates. The criteria are: AIC (Akaike Information Criterion), BIC (Bayesian Information Criterion), CAIC (Consistent Akaike Information Criterion) and HQIC (Hannan-Quinn Information Criterion). These statistics are given by $AIC=-2l(\hat{\alpha})+2p$, $BIC=-2 l(\hat{\alpha})+ 2\ln(n)$, $CAIC=-2l(\hat{\alpha})+ p(\ln (n)+1)$ and $HQIC=-2l(\hat{\alpha})+2p\ln(\ln(n))$, where $l(\hat{\alpha})$ denotes the log-likelihood function evaluated at the MLEs, $p$ is the number of model parameters and $n$ is the sample size. The model with lowest values for these statistics could be chosen as the best model to fit the data. The values of  MLE of the parameter, $l(\hat{\alpha})$, AIC, BIC, HQIC, CAIC \& K-S Statistic are displayed in Table $3$. Among all other competitive models, it is to be noted that the IXGD($\alpha$) has the lowest values of $l(\hat{\alpha})$, AIC, BIC, HQIC, CAIC \& K-S and so it could be chosen as the best model to fit the given data sets.
\begin{itemize}
\item {\bf Data Set I:} Postate cancer data taken from Collett ($2003$) of size $38$. For this data, we have compared the proposed model IXGD with well known one parameter inverted family of distributions, viz., inverted Exponential distribution (IED), inverse Lindley distribution (ILD), inverse Rayleigh distribution (IRD) and found that IXGD is the better choice for the considered data set [see, Table $3$].\\

\item {\bf Data Set II:} Item failure data represents the $46$ repair times (in hours) have taken from Chhikara and Folks ($1997$), initially considered by Chhikara and Folks ($1977$), for an airborne communication transceiver. Here also, we have compared the proposed model IXGD with well known one parameter inverted family of distributions, viz., inverted Exponential distribution (IED), inverse Lindley distribution (ILD), inverse Rayleigh distribution (IRD) and found that IXGD is the better choice for the considered data set [see, Table $3$].\\
 \begin{table}[ht]
  \textbf{Table 3: The model fitting summary for the considered data sets based on MLE.}\\ 
  \begin{tiny}
  \begin{center}
  \begin{tabular}{|l|l|l|l|l|l|l|l|l|l|l|l|l|l|}\hline
 Data    & Model  &  Estimate       &  Negative           &  AIC       & BIC  & CAIC & HQIC  & K-S        \\
  Set    &        & of $\theta$     &  Log-likelihood     &            &             & &   &  Statistic \\
  \hline
  \hline
  I         & IED     & $24.97312$   & $200.1996$    & $402.3996$ & $404.0369$   & $405.0369$  &$402.9819$ & $0.31005$ \\
 \hline
           & ILD      & $25.90154$   & $200.2675$    & $402.5350$ & $404.1726$    &$405.1726$ & $403.1176$ & $0.29978$ \\
 \hline 
           & IRD      & $120.6323$   & $256.2503$    & $514.5006$ & $516.1382$    &$517.1382$& $515.0832$ & $0.67427$ \\
  \hline
            & IXGD    & $26.82069$   & $199.4590$    & $400.9181$ & $402.5570$    & $403.5570$&$401.5007$  & $0.30710$ \\
  \hline 
  &&&&&&&&\\
  \hline
   II         & IED     & $1.13620$   & $100.6971$   & $203.3941$ & $205.2228$  &$206.2281$& $204.0792$ & $0.06790$  \\
   \hline
            & ILD      & $1.57712$   & $100.1692$    & $204.3385$ & $206.1671$  &$207.1671$& $205.0235$ & $0.08128$  \\
   \hline 
            & IRD      & $0.60590$   & $128.0239$    & $258.0478$ & $259.8763$  &$260.8765$& $258.7329$ & $0.33770$  \\
    \hline
             & IXGD    & $1.90130$   & $101.1312$    & $204.2623$ & $206.0909$  &$207.0909$& $204.9473$ & $0.06720$  \\
    \hline 
  \end{tabular}
  \end{center}
  \label{tab3}
  \end{tiny}
  \end{table}
Also, the estimates of the parameter $\theta$ and reliability characteristics of IXGD for both the data sets are computed using different methods of estimation [see, Table $4$]. 
\begin{table}[ht] 
\textbf{Table 4: Widths of ACI of the parameter $\theta$ for IXGD}\\
\begin{center}
\begin{tabular}{|l|l|l|l|l|l|l|l|l|l|l|l|l|}\hline
Data Set  & Model  & $\hat{\theta}_{MLE}$  & $\hat{\theta}_{LSE}$ & $\hat{\theta}_{WLSE}$ & $\hat{\theta}_{CME}$ & $\hat{\theta}_{MPSE}$ & ACI \\
          &         &                      &                      &                       &                      &                       &     \\
\hline
I         & IXGD    & $26.82069$           &  $33.22330$          &  $26.63840$           &  $33.19222$          &  $26.75462$           & $0.5346$ \\
\hline
II        & IXGD    & $01.90130$          &  $01.98668$            &  $01.95813$            &  $01.99161$        &  $01.88956$           & $0.8752$ \\
\hline
\end{tabular}
\end{center}
\label{tab4}
\end{table}
 \end{itemize}

\section{Concluding Remarks.}
In this article, we have proposed a new probability model, namely, IXGD by considering the inverse of XGD. Different statistical characteristics and properties have been discussed. Different methods of estimation have been discussed for estimating the unknown parameter of the proposed model. The comparison among the considered estimators of $\theta$ have been carried out using Monte Carlo simulation study and it has been noticed that MPSE of $\theta$ performed better in MSE sense. Further, $95\%$ ACI of $\theta$ has been calculated using MPSE of $\theta$ as it performed better among the other estimators. Finally, two real data sets have been analyzed for illustration purposes of the study.  
\section*{References}
\begin{enumerate}
\item Cheng R. C. H., Amin N. A. K. (1979). Maximum product-of-spacings estimation with applications to the log-normal distribution, University of Wales IST, Math Report, 79-1.

\item  Cheng, R. C. H.  and Amin, N.A.K. (1983). Estimating parameters in continuous univariate distributions with a shifted origin. {\it Journal of the Royal Statistical Society: Series B Statistical Methodology}, {\bf 3}, 394-403.

\item Chhikara, R. S. and Folks, J. L. (1977). The inverse gaussian distribution as a lifetime model. {\it Technometrics}, {\bf 19}, 461-468.

\item Collett (2003). Modelling Survival Data in Medical Research. 2, Chapman and Hall/CRC.

\item Ghosh, S. R. Jammalamadaka, (2001). A general estimation method using spacings. {\it Journal of Statistical Planning and Inference}, 93.

\item Ihaka, R. and Gentleman, R. (1996). R: A language for data analysis and graphics. {\it Journal of Computational and Graphical Statistics}, {\bf 5}, 299-314.

\item  Keller, A. Z and Kamath, A. R (1982). Reliability analysis of CNC Machine tools. {\it Reliability Engineering}, {\bf 3}, 449-473. http://dx.doi.org/10.1016/0143-8174(82)90036-1.

\item MacDonald, P. D. M. (1971). Comment on an estimation procedure for mixtures of distributions by Choi and Bulgren. {\it Journal of Royal Statistical Society: Series B}, {\bf 33(2)}, 326-329.

\item Ranneby, B. (1984). The maximum spacing method. an estimation method related to the maximum likelihood Method. {\it Scandinavian Journal of Statistics}, {\bf 11(2)}, 93-112.

\item Sen, S., Maiti, S. S. and Chandra, N. (2016). The xgamma Distribution: statistical properties and Application {\it Journal of Modern Applied Statistical Methods}, {\bf 15(1)}, 774-788.

\item Shaked, M., and Shanthikumar, J. G. (1994). Stochastic orders and their applications. New York, NY: Academic Press.

\item Sharma, V. K., Singh, S. K., Singh, U. and Agarwal, V. (2015). The inverse Lindley distribution: a stress-strength reliability model with application to head and neck cancer data. {\it Journal of Industrial and Production Engineering}, {\bf 32(3)}, 162-173.

\item Swain, J., Venkatraman, S. and  Wilson, J. (1988). Least squares estimation of distribution function in Johnsons translation system. {\it Journal of Statistical Computation and Simulation}, {\bf 29}, 271-297.

\item Voda, V. G. (1972). On the inverse Rayleigh random variable. {\it Rep. Stat. Appl. Res. Jues}, {\bf 19(4)}, 13-21.
\end{enumerate}
\end{document}